\documentclass[aps,prb,twocolumn,showpacs,preprintnumbers,amsmath,amssymb]{revtex4}
\usepackage{graphicx}
\usepackage{dcolumn}
\usepackage{bm}
\usepackage{hyperref}

\begin{document}

\preprint{AC-ZTFS-4-2009}

\title{Interorbital pair scattering in clean and impure superconductors}

\author{Anna Ciechan}
\author{Karol Izydor Wysoki\'nski}%
 \email{karol@tytan.umcs.lublin.pl}
\affiliation{%
Institute of Physics and Nanotechnology Centre, 
M. Curie-Sk\l{}odowska University,\\ 
 ul. Radziszewskiego 10, Pl 20-031 Lublin, Poland
}%
 
\date{\today}

\begin{abstract}
The calculations of the local and global properties of
two band superconductors have been presented with  
particular attention to the role of the inter-orbital 
scattering of pairs. The properties of such superconductors  
are very different from a single band or 
typical two band systems with dominant intra-band pairing interactions. The role
of Van Hove singularity in one of the bands on the properties 
of intra-band  clean superconductor has been discussed.
It leads to marked increase of superconducting transition temperature 
in the weak coupling limit.
We study  the  inhomogeneous
systems in which the characteristics change from place to place
by solving the Bogolubov-de Gennes equations
for small clusters. The suppression of the superconducting
order parameter by the single impurity   
scattering the fermions between bands is contrasted with 
that due to intra-band impurity 
scattering. The results obtained for impure systems have been shown 
as a maps of local density of states,
 the order parameter and gap function. They can 
 be directly compared with STM spectra of the real material.
\end{abstract}

\maketitle

\section{\label{sec:level1}Introduction}
Already in the fifties and sixties the main properties of the two band 
superconductors have been clarified \cite{Suhl1959,moskalenko1959,Suffczynski1962,Kondo1963}. 
At that time, however,
the existing materials did not show clear evidence of two band behavior.
The experimental situation has changed with the discovery of the high
temperature superconducting oxides \cite{bednorz1986} and even more
with subsequent discoveries of strontium ruthenate \cite{maeno1994}, magnesium 
diboride \cite{nagamatsu2001}  and iron pnictides \cite{kam1,kam2}. 
Even though all of these systems have a number of bands in the vicinity
of the Fermi energy their presence  shows up in quite a different way.

Magnesium diboride clearly shows two different gaps of the same
symmetry \cite{wang2001,angst2002}. In strontium ruthenate 
the three band model seems to be necessary
to explain its puzzling properties \cite{mackenzie2003,wysokinski2003}. 
The model of superconductivity in the iron pnictides is a matter of
ongoing debate \cite{mazin2009,sadovski2008,ivanovski2008, izyumov2008}. 
The iron pnictides possess a large number of bands around
 the Fermi  energy and few of them seem to play important role
in the superconducting state \cite{ding2008,chubukov2008,kuroki2008}. 
Two band model has been proposed as 
 minimal model of these superconductors \cite{raghu2008}. 

With two bands near the Fermi energy one generally 
expects formation of intra-band and inter-band pairs. 
In the later case the pairs have in general non-zero 
center of mass momentum \cite{liu2003}. The simpler case \cite{Suhl1959}  
of superconductivity with intraband pairs, which can be scattered
between two bands seem to be relevant in modelling of pnictides.
Indeed, there are strong theoretical \cite{fanfarillo2009,laad2009} 
arguments that inter-band interactions may be important 
in these system. These findings make pnictides
different from MgB$_2$, in which main coupling mechanism 
is intraband \cite{liu2001}. Thus the detailed study of 
inter-orbital \cite{orbital} pair scattering mechanism of superconductivity 
is timely and of importance.  The issue has recently been discussed 
in connection with both cuprate \cite{ord2000} and 
pnictide superconductors \cite{dolgov2009}.

In this paper we are mainly interested
in the properties of inter-orbital only mechanism of superconductivity.
We shall study both clean homogeneous and impure models.
Without loss of generality we shall denote two orbitals as 1 and 2.
The interaction $U_{11}$ ($U_{22}$) is responsible for superconducting 
instability inside a band formed by orbitals 1 (2), while 
 $U_{12}$ promotes the scattering of superconducting 
pair between orbitals 1 and 2. The impurity scattering potential 
is assumed in general form 
$\sim V_{imp}^{\lambda\lambda '}c_{i\lambda\sigma}^+c_{i\lambda '\sigma}$.
It scatters electrons from site $i$, orbital $\lambda'$ into  
orbital $\lambda$ of the same site.
If $\lambda=\lambda'$ we call such impurities  intra-band, 
if $\lambda\ne\lambda'$
  inter-band. 

The organisation of the rest of the paper is as follows. 
Section 2 presents the general Hamiltonian of
the two orbital model and the Bogolubov - de Gennes (BdG) approach used
to solve it. The homogeneous superconductors are discussed in Section 3, where 
we study {\it inter alia} the effect of Van Hove singularity in the density of
states in one of the bands on the properties of the superconductors
with inter-band pair scattering only. The changes induced in the superconductor
by single intra-band or inter-band impurity are discussed in Section 4, while
the finite concentration of impurities is considered in Section 5. We end up with
the discussion of our results and their relevance to most
prominent two band superconductors: MgB$_2$ and iron pnictides.

\section{Hamiltonian for the two orbital superconductor}
We start with general Hamiltonian in a real space describing 
the system with two orbitals. We assume the spin-independent 
effective pairing interaction between fermions 
in various  orbital states.  The randomness
in the system is easily incorporated $via$ site dependence of parameters.  
 The Hamiltonian reads 
\begin{eqnarray}
H&=&\sum_{ij,\lambda\lambda',\sigma}(-t_{ij}^{\lambda\lambda'}+V_{imp}^{\lambda\lambda'}(\vec{r_i})\delta_{ij})
c^+_{i\lambda\sigma}c_{j\lambda'\sigma}
\nonumber
\\
&+&
\sum_{i,\lambda,\sigma}(e_{\lambda}-\mu)c^+_{i\lambda\sigma}c_{i\lambda\sigma}
\nonumber
\\
&+&\sum_{i,\lambda_1\lambda_2,\lambda_3\lambda_4}U_{\lambda_1\lambda_2\lambda_3\lambda_4}(\vec{r_i})
c^+_{i\lambda_1\uparrow}c^+_{i\lambda_2\downarrow}c_{i\lambda_3\downarrow}
c_{i\lambda_4\uparrow},
\label{h1}
\end{eqnarray}
where $c_{i\lambda \sigma}^{+}$, $c_{i\lambda \sigma}$ are creation and annihilation 
operators of electrons with spin $\sigma=\uparrow,\downarrow$ at the lattice site  
$\vec{r_i}=i$ in the orbital $\lambda$. 
$\epsilon_{\lambda}$ is the electron energy and $\mu$ is the chemical potential.
$t_{ij}^{\lambda \lambda '}$ are the hopping integrals between the same or different
orbitals (if $\lambda\ne\lambda '$).
$U_{\lambda_1 \lambda_2\lambda_3\lambda_4}(\vec{r_i})$  denotes  interactions, 
which are attractive for $U_{\lambda_1 \lambda_2\lambda_3\lambda_4}(\vec{r_i})<0$. The dependence
of the interaction parameters on the position $\vec{r_i}$ allows to treat 
systems with inhomogeneous pairing. 

We use standard mean-field decoupling valid for a spin singlet superconductor
and get the following effective Hamiltonian
\begin{eqnarray}
H^{MFA} &=&
\sum_{ij,\lambda\lambda ',\sigma}(-t_{ij}^{\lambda\lambda '}+V_{imp}^{\lambda\lambda'}(\vec{r_i})\delta_{ij})
 c^+_{i\lambda\sigma}c_{j\lambda '\sigma}
\nonumber
\\
&+&\sum_{i,\lambda ,\sigma}(\epsilon _\lambda+V_{\lambda,\sigma}(\vec{r_i})-\mu)
c_{i\lambda\sigma}^+c_{i\lambda\sigma}
\nonumber\\
&+&\sum_{i,\lambda\lambda'}\left( 
\Delta_{\lambda\lambda'}(\vec{r_i})c^+_{i\lambda\uparrow}c^+_{i\lambda'\downarrow}+
h.c.
\right),
\label{h-MFA}
\end{eqnarray}
where the order parameters $\Delta_{\lambda\lambda '}(\vec{r_i})$ are related to the pairing 
correlation functions $f_{\lambda\lambda '}(\vec{r_i})=<c_{i\lambda\downarrow}c_{i\lambda'\uparrow}>$
through 
\begin{eqnarray}
\Delta_{\lambda_1\lambda_2}(\vec{r_i})=
-\sum_{\lambda_3\lambda_4}U_{\lambda_1\lambda_2\lambda_3\lambda_4}(\vec{r_i})f_{\lambda_3\lambda_4}(\vec{r_i}).
\label{del}
\end{eqnarray}
The local Hartree terms $V_\lambda(\vec{r_i})$ depend on the  number of particles at given site $n_{\lambda\sigma}(\vec{r_i})=
<c^+_{i\lambda\sigma}c_{i\lambda\sigma}>$ 

\begin{eqnarray}
V_{\lambda\sigma}(\vec{r_i})=
\sum_{\lambda '}U_{\lambda'\lambda\lambda\lambda'}(\vec{r_i})n_{\lambda' -\sigma}(\vec{r_i}).
\label{HT}
\end{eqnarray}
We consider here only diagonal correlations 
$<c^+_{i\lambda\sigma} c_{i\lambda'\sigma'}>=\delta_{\lambda\lambda '}\delta_{\sigma\sigma '}n_{\lambda\sigma}(\vec{r_i})$.

The Hamiltonian (\ref{h-MFA}) is diagonalised with help of the 
Bogolubov - Valatin transformation \cite{Bogoliubov, Valatin}
\begin{eqnarray}
c_{i\lambda\uparrow}=\sum_\nu\left(u_{\lambda\nu}(\vec{r_i})\gamma_{\nu\uparrow}-
v^*_{\lambda\nu}(\vec{r_i})\gamma^+_{\nu\downarrow}
\right),\\
c_{i\lambda\downarrow}=\sum_\nu\left(u_{\lambda\nu}(\vec{r_i})\gamma_{\nu\downarrow}+
v^*_{\lambda\nu}(\vec{r_i})\gamma^+_{\nu\uparrow}
\right)
\label{BVT}
\end{eqnarray}
leading to the Bogolubov- de Gennes (BdG) equations for amplitudes 
 $u_{\lambda\nu}(\vec{r_i})$, $v_{\lambda\nu}(\vec{r_i})$ and eigenenergies $E_\nu$ 
\begin{eqnarray}
&\sum_{j,\lambda'}K_{ij}^{\lambda\lambda '}
u_{\lambda'\nu}(r_j)
+\sum_{\lambda'}\Delta_{\lambda\lambda'}(\vec{r_i})v_{\lambda'\nu}(\vec{r_i})=&
\nonumber
\\
&
E_\nu u_{\lambda\nu}(\vec{r_i}),&
\label{BdGEa}
\\
&-\sum_{j,\lambda'}K_{ij}^{\lambda\lambda '}
v_{\lambda'\nu}(r_j)
+\sum_{\lambda'}\Delta^*_{\lambda\lambda'}(\vec{r_i})u_{\lambda'\nu}(\vec{r_i})=&
\nonumber
\\
&
E_\nu v_{\lambda\nu}(\vec{r_i}),&
\label{BdGEb}
\end{eqnarray}
where the operator $K_{ij}^{\lambda\lambda '}$ reads 
\begin{eqnarray}
K_{ij}^{\lambda\lambda '}=
(e_\lambda-\mu+V_{\lambda\sigma}(\vec{r_i}))\delta_{ij}\delta_{\lambda\lambda'}
+V_{imp}^{\lambda\lambda '}(\vec{r_i})\delta_{ij}
-t^{\lambda\lambda'}_{ij}.
\label{Kij}
\end{eqnarray}

The pairing parameters $\Delta_{\lambda\lambda '}(\vec{r_i})$ and Hartree potentials $V_\lambda(\vec{r_i})$ 
are in turn expressed  in terms of eigenfunctions and eigenergies $u_{\lambda\nu}(\vec{r_i})$, 
$v_{\lambda\nu}(\vec{r_i})$, $E_{\nu}$ as \cite{ketterson1999}
\begin{eqnarray}
n_{\lambda}(\vec{r_i})=
\sum_\nu\left(
|u_{\lambda\nu}(\vec{r_i})|^2f_\nu+
|v_{\lambda\nu}(\vec{r_i})|^2(1-f_\nu)
\right),
\label{NP}
\end{eqnarray}

\begin{eqnarray}
f_{\lambda\lambda '}(\vec{r_i})&=&
\sum_\nu[
u_{\lambda\nu}(\vec{r_i})v^*_{\lambda '\nu}(\vec{r_i})(1-f_\nu) 
\nonumber\\
&-&
u_{\lambda '\nu}(\vec{r_i})v^*_{\lambda\nu}(\vec{r_i})f_\nu].
\label{PCF}
\end{eqnarray}
In the above formulae $f_\nu=\left(e^{E_\nu/k_BT}+1\right)^{-1}$ denotes the  
Fermi-Dirac distribution function 
of quasi-particles. The total number of particle in the given band (to be denoted 
by the same index as
the orbital) is given by $N_\lambda=\sum_i n_\lambda (\vec{r_i})$.

The local density of states (LDOS) $N(\vec{r_i},E)$ is directly accessible in scanning tunneling 
microscope (STM) measurements and is proportional to the local conductance $dI(\vec{r_i},V)/dV$.
In the two band system it is a sum of local  densities of states 
of the individual bands $ N(\lambda,\vec{r_i},E) $  
\begin{eqnarray}
N(\lambda,\vec{r_i},E)&=&
\sum_\nu [|u_{\nu\lambda}(\vec{r_i})|^2\delta(E-E_{\nu})
\nonumber
\\
&+&|v_{\nu\lambda}(\vec{r_i})|^2\delta(E+E_{\nu})].
\label{LDOS}
\end{eqnarray}
Obviously we have at each site $ N(\vec{r_i},E) = N(1,\vec{r_i},E) + N(2,\vec{r_i},E) $.
 For a clean system equations (\ref{BdGEa},\ref{BdGEb}) can be Fourier transformed
 and written (in closely analogous form) in reciprocal space. For the 
 impure systems with broken translational symmetry the Bogolubov - de Gennes
 equations (\ref{BdGEa},\ref{BdGEb}) are solve self-consistently 
 in real space for a small n$\times$m cluster with periodic boundary conditions.
 For a two orbital model the typical size of the cluster is 20$\times$30.
 In the next section we start with the comparison of our real space
(for small cluster) calculations with (numerically) exact results obtained in reciprocal space
 ($i.e.$ for bulk system). 

\section{Homogeneous superconductors}

In this section we shall discuss some properties of homogeneous 
two band superconductors paying special attention to the comparison of the
accuracy of small cluster calculations with bulk system. We also consider the effect 
of Van Hove singularity in one of the bands on the properties 
of inter-band pairing superconductivity and the role of various 
inter-band couplings.

\subsection{Small clusters {\it vs.} bulk systems}

We start with the homogeneous system with two orbitals denoted 1 and 2.
The superconductor is described by the following set of parameters.
The inter-band interaction \cite{Suhl1959} has the form of pair scattering 
only $U_{12}=U_{1122}$, while
two intra-band interactions are $U_{11}=U_{1111}$  and $U_{22}=U_{2222}$. 
We consider two-dimensional square lattice with non-zero hopping integrals 
between the nearest neighbor sites only 
$t_\lambda=t^{\lambda\lambda}_{ij}$ and hybridization 
$t_{12}=t^{12}_{ij}$. We set the direct 
hoping between orbitals no. 1 as our energy unit $t_1=t=1$.
 Since we are ignoring  the possibility of  inter-orbital pairs, 
we use the simpler notation $\Delta_1=\Delta_{11}$ and $\Delta_2=\Delta_{22}$.

\begin{figure}
\rotatebox{-90}{\scalebox{0.3}{
\includegraphics{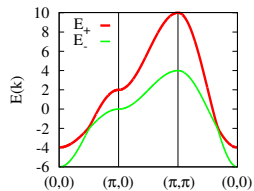}}}
\caption{\label{bands} Energy bands of the two dimensional square
lattice with parameters $e_2-e_1=2t, t_1=t,t_2=2t,t_{12}=0.05t$.
The chemical potential is located at $\mu=0$ and 
the resulting total number of particles $n=1.62$.}
\end{figure}

Fig. (\ref{bands}) shows the single particle energy bands 
along main symmetry direction in the two dimensional Brillouin 
zone obtained for the following set of 
parameters $e_2-e_1=2t, t_1=t,t_2=2t,t_{12}=0.05t$. The chemical potential 
$\mu=0$ and the total number of carriers $n=1.62$. 

\begin{figure}
\rotatebox{-90}{\scalebox{0.2}{
\includegraphics{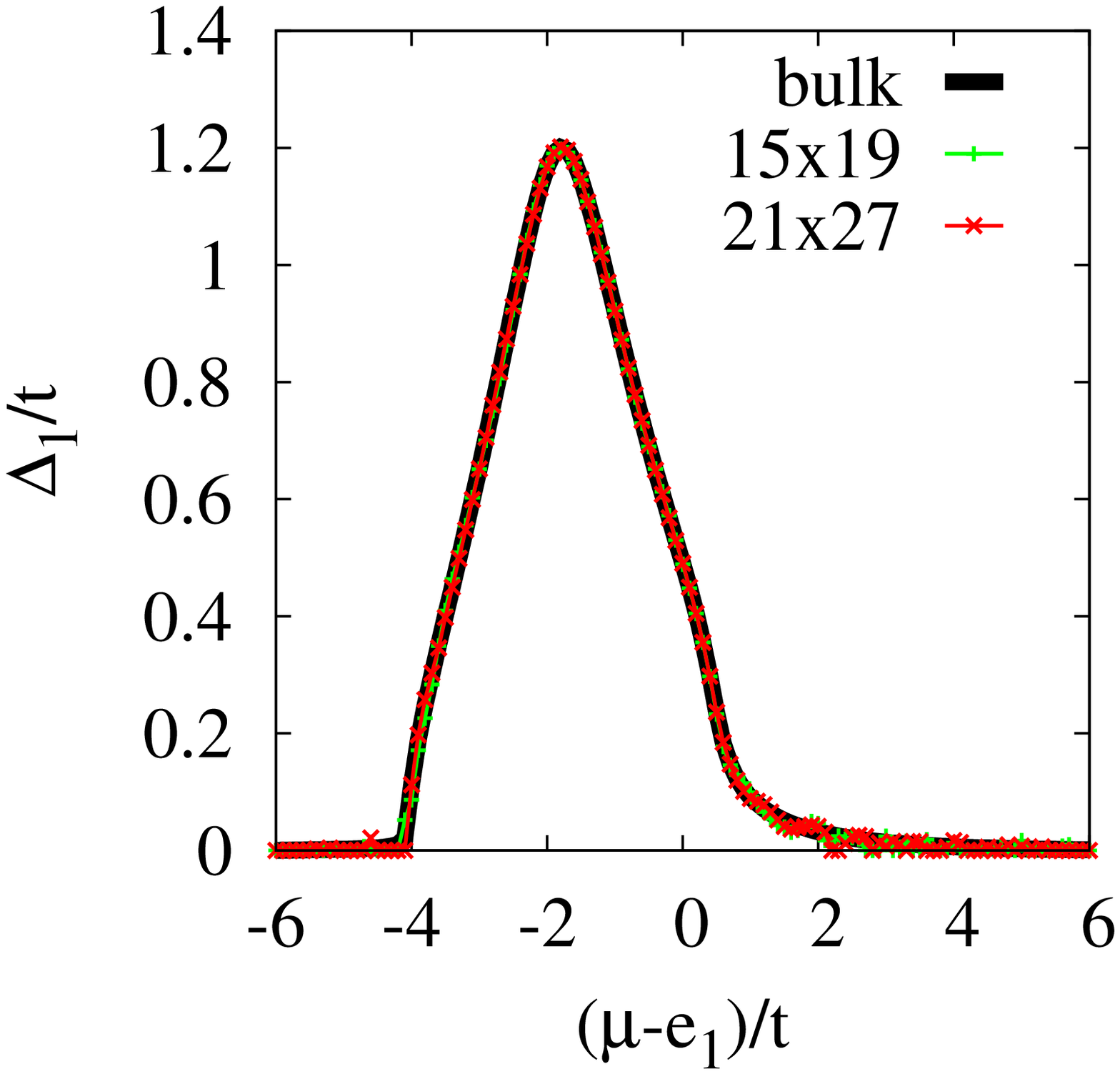}}}
\rotatebox{-90}{\scalebox{0.2}{
\includegraphics{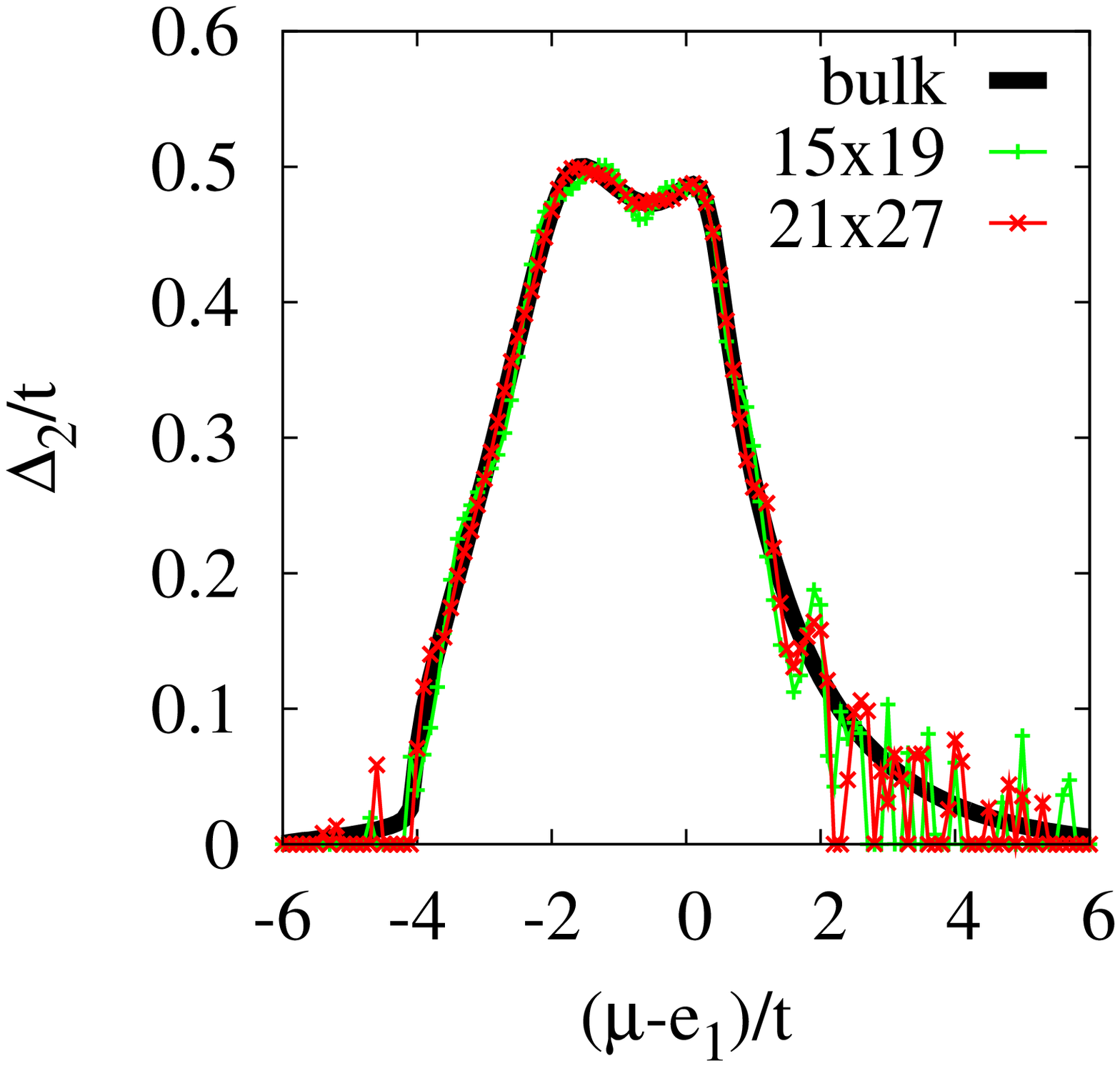}}}
\rotatebox{-90}{\scalebox{0.2}{
\includegraphics{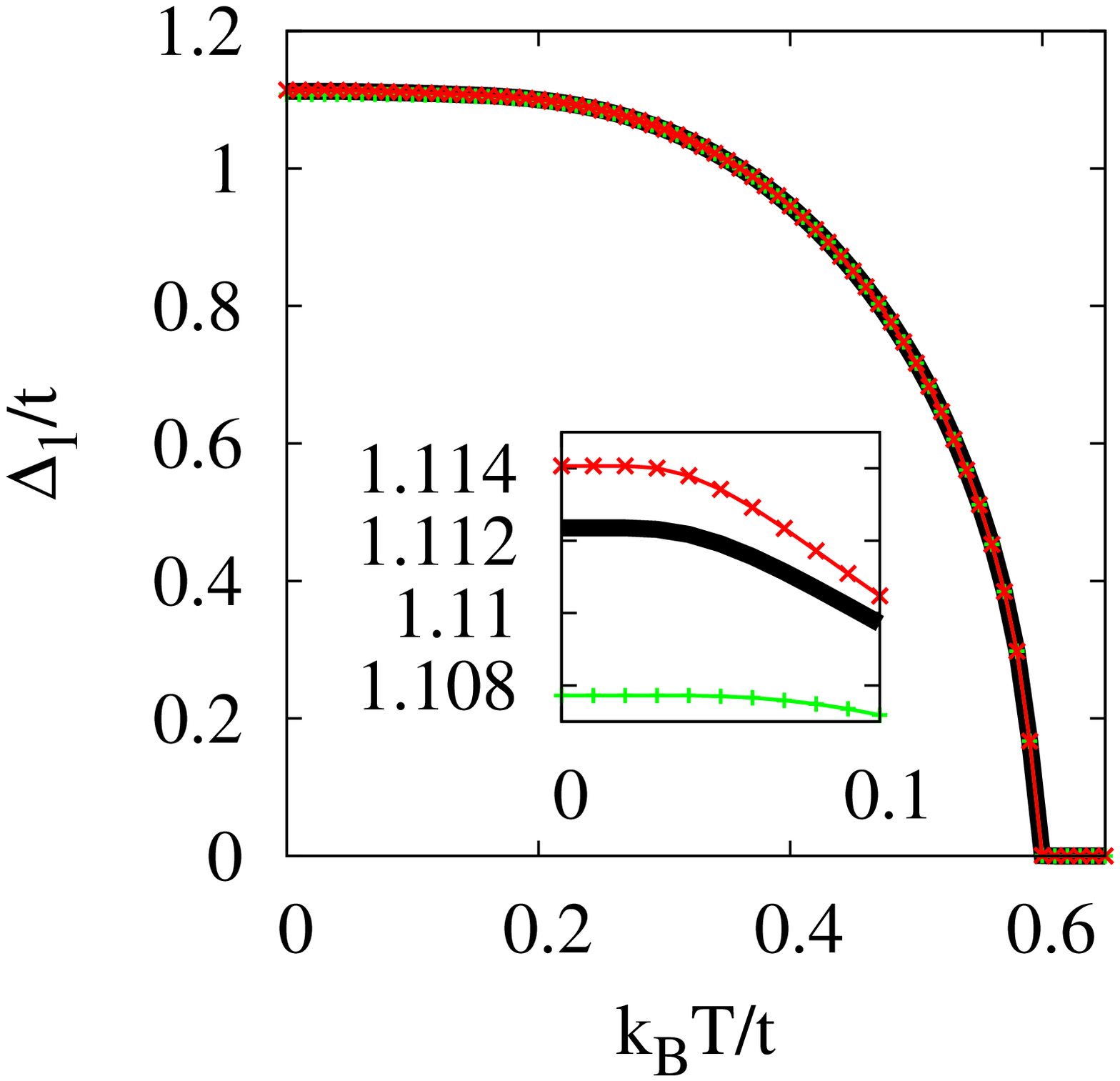}}}
\rotatebox{-90}{\scalebox{0.2}{
\includegraphics{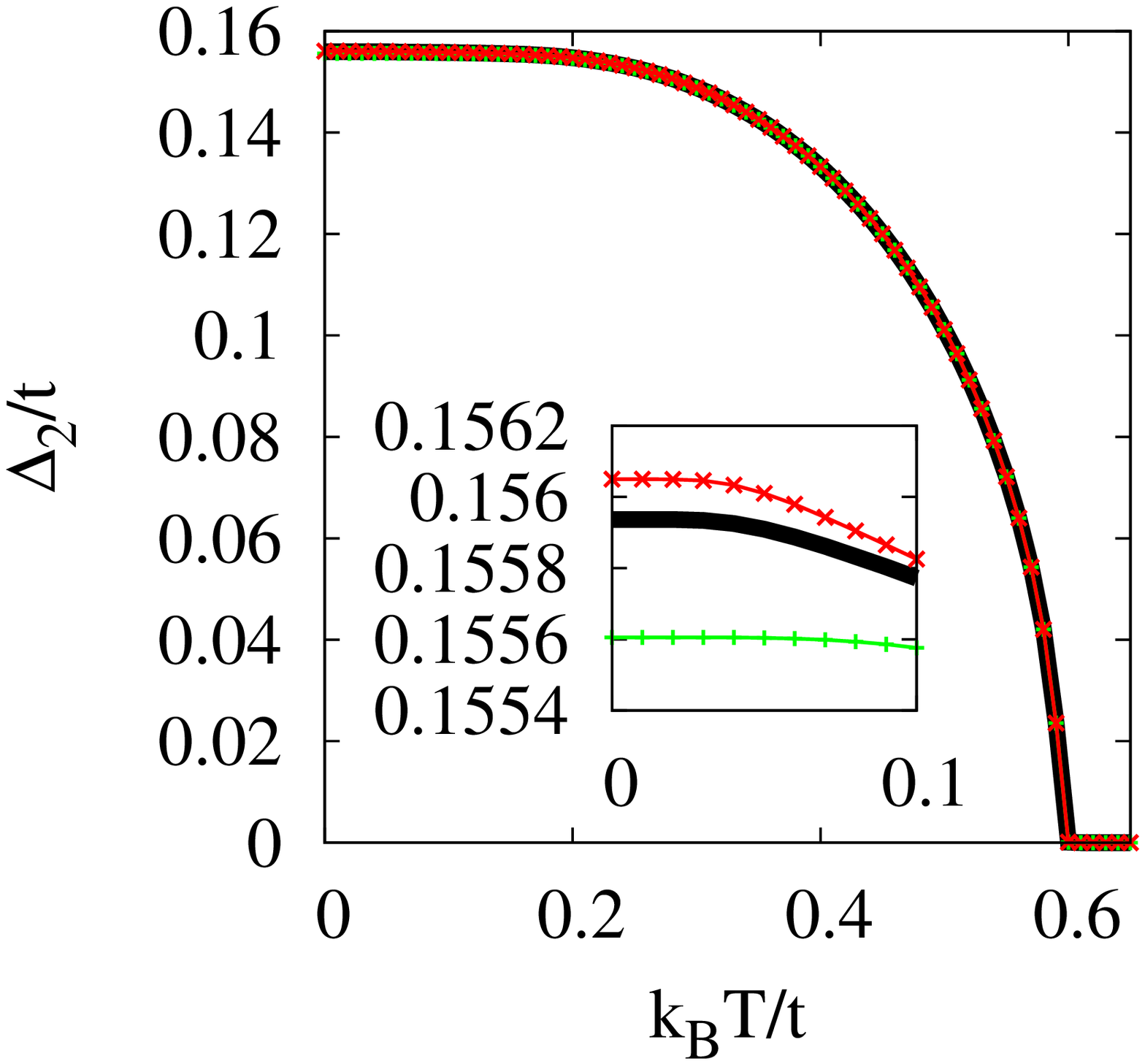}}}
\caption{\label{k-r} 
The upper panels show the dependence of the $\Delta_{\lambda}$ for 
both bands on the chemical potential $\mu$ for bulk system (solid curve)
and for clusters of size 15$\times$19 (thin line with plusses) and 21$\times$27 
(line with crosses). The other parameters    are: 
$e_2-e_1=2t, t_1=t,t_2=2t,t_{12}=0.05t$, $U_1=-3.5t$, $U_2=-3t$, $U_{12}=-0.5t$. 
At the lower panels temperature dependence  of $\Delta_{\lambda}$
are shown for $e_1-\mu=2t$.  The insets show 
the  data near T=0 on the expanded scale. }
\end{figure}

Fig. (\ref{k-r}) compares the solutions obtained for the bulk system
with those for small clusters of various size. We consider here
the bulk data as exact. The accuracy of determination of the 
gap parameter for bulk homogeneous system 
(in our case assumed to be of the order of 10$^{-6}$t) 
is only limited by the time of calculations. 
We have found that at the band center the results obtained for
clusters with size greater than 400 sites are acceptable.
Relative changes of the gaps with respect to the bulk values
 $$\delta\Delta_{\lambda}=(\Delta_{\lambda}(L)-
\Delta_{\lambda}^{bulk})/\Delta_{\lambda}^{bulk} \cdot 100\%$$
are in the range 
of $\delta\Delta_{1}<0.15$\% in the first band and slightly greater
 $\delta\Delta_{2}<1.5$\% for the second band. Well inside the bands
 the spectrum is quasi-continuous and the results agree very well  
with bulk data but near the band edges 
 the spectrum of finite clusters is discrete and the differences are 
 larger (c.f. Fig. (\ref{k-r})).
 
\subsection{Inter-orbital pairing only superconductor - the role of Van Hove singularity}
As a general rule one finds that the inter-orbital scattering $U_{12}$
plays a minor role in superconductors with dominant intra-band   
interactions \cite{wysokinski2009}. This interaction, however
couples two bands and may lead to the increase of the superconducting 
transition temperature \cite{Kondo1963,bussmann-holder2004}. 
The situation changes drastically if the inter-band
pairing is the only existing interaction. The properties of
the superconductors with dominant inter-band scattering are markedly different   
from those with dominant intra-band interactions. In particular \cite{mazin2008}, 
the superconducting 
transition takes place for arbitrary sign of $U_{12}$. The value of the gap
in the first band is determined by the interaction $U_{12}$ and the density 
of states in the second band and {\it vice versa}, the gap in the second band 
is proportional to the partial density of states (DOS) at the Fermi level in the first band.

This can easily be seen from the general two band BCS equations \cite{Suhl1959} 
\begin{equation}
\Delta_{1}(1+U_{11}F_1)=-U_{12}\Delta_{2}F_2,\\
\nonumber
\end{equation}
\begin{equation}
\Delta_{2}(1+U_{22}F_2)=-U_{12}\Delta_{1}F_1,
\end{equation}
where
\begin{equation}
F_\lambda=\int_0^{\hslash\omega_c} dE N_{\lambda}(E)\frac{\tanh{\sqrt{E^2+\Delta_{\lambda}^2} 
\over 2k_BT}}{\sqrt{E^2+\Delta_{\lambda}^2}}
\end{equation}
and $N_{\lambda}(E)$ denotes single particle density of states
in the band $\lambda$.  In this discussion we  interested in the limit of inter-band 
pair scattering  only. For $U_{11}=U_{22}=0$  the above equations reduce to

\begin{equation}
\Delta_{1}=-U_{12}\Delta_{2}\int_0^{\hslash\omega_c} dE N_{2}(E)\frac{\tanh{\sqrt{E^2+\Delta_{2}^2} 
\over 2k_BT}}{\sqrt{E^2+\Delta_{2}^2}},
\nonumber
\end{equation}
\begin{equation}
\Delta_{2} =-U_{12}\Delta_{1}\int_0^{\hslash\omega_c} dE N_{1}(E)\frac{\tanh{\sqrt{E^2+\Delta_{1}^2} 
\over 2k_BT}}{\sqrt{E^2+\Delta_{1}^2}}.
\label{intrab}
\end{equation} 
It is clear from equations (\ref{intrab}) that the value 
of $\Delta$ in the second band is determined by the
density of states in the first one and {\it vice versa}. It is also obvious
that the nonzero solutions can be obtained for both signs and arbitrary 
small value  of the coupling $U_{12}$. 
For positive value of it the order parameters in the two bands have oposite signs, while for
negative $U_{12}$ they are of the same sign.  The inter-orbital pairing only model
has a number of unusual features.  
It has been found  \cite{dolgov2009,bang2008} that 
the ratio $\Delta_{2}/\Delta_{1}=\sqrt{N_1/N_2}$, where $N_2(N_1)$ is
the density of states in band 2(1) at the Fermi level and 
the superconductiong transition temperature
of the system is given by the BCS-like expression 
\begin{equation}
T_c=1.136{\hslash\omega_c \over k_B} \exp\left({-1 \over\lambda_{eff}}\right)
\label{bcs-tc}
\end{equation}
with $\lambda_{eff}=\lambda_0=\sqrt{U_{12}^2N_1N_2}$.  
  
It often happens that the Fermi level in  superconductors
lies close to the Van Hove singularity. In layered systems with
nesting properties of the (quasi - two dimensional) Fermi surface the
density of states near the Van Hove singularity changes  logarithmically
\begin{equation}
N(E)=N_0\ln(2W/|E|) \Theta(|E|-W)
\end{equation}
with 2W being the band width and $\Theta(x)$ the step function. 
It is known that the existence of such  
singularity modifies \cite{markiewicz1997} the BCS expression 
for the transition temperature (\ref{bcs-tc}).
In particular, in the one band case and the weak coupling 
limit \cite{labbe1987}  $\lambda \ll 1$ it leads to increase of the
superconducting transition by changing the effective interaction: 
${1 \over \lambda_{eff}} = \sqrt{2 \over \lambda}$.

Here we assume  the density of states in the second band 
to be singular $N(2,E)=N_2\ln(2W/|E|)$ near the Fermi energy,
while that of the first band  flat $N(1,E)=N_1$. 
Near $T_c$ equations (\ref{intrab}) are linearised,  we approximate 
$\tanh(x)=min(x,1)$ and find (we use  here $\hslash=k_B=1$)
\begin{eqnarray}
\Delta_{1}&=&-\Delta_{2} U_{12}N_2[1+\ln{2W \over\omega_c}+\ln{\omega_c \over 2T_c}  \nonumber  \\
 &+&\ln{2W \over\omega_c}\ln{\omega_c \over 2T_c} 
 +{1\over2}(\ln{\omega_c \over 2T_c})^2 ]   \nonumber \\
 \Delta_{2}&=&-\Delta_{1} U_{12}N_1[1+\ln{\omega_c \over 2T_c}].
\end{eqnarray}
 The analysis of the above set of equations in the weak
coupling limit ($U_{12}\rightarrow 0$) leads to the
 approximate BCS like expression for the superconducting transitions temperature
 with  ${1\over \lambda_{eff}}= {2^{1/3} \over \lambda_{0}^{2/3}}=({2\over U_{12}^2N_1N_2})^{1/3}$ 
 and to the modification of the 
 prefactor, which changes from $1.136\omega_c$ to $1.136\omega_c(2W/\omega_c)^{2/3}$. 
 It is interesting to note that up to the prefactor  the Van Hove singularity 
 in one of the bands increases $T_c$ of intra-band superconductor at the very weak coupling only
 $\lambda_0\ll 1$.  

In a similar way one can calculate the effect of Van Hove
singularity on the ratio of the gaps $\Delta_2\over\Delta_1$ at zero temperature. 
One finds 
\begin{eqnarray}
\Delta_{1}(0)&=&-\Delta_{2}(0) U_{12}N_2[1-\ln{\Delta_{2}(0)\over 2W }-{1\over2}(\ln{\omega_c \over 2W})^2 \nonumber \\ 
 &+&{1\over2}(\ln{\Delta_{2}(0)\over 2W})^2] \nonumber \\
\Delta_{2}(0)&=&-\Delta_{1}(0) U_{12}N_1\ln{2\omega_c \over \Delta_{1}(0)}
\end{eqnarray}
Even  in the extreme   weak coupling limit $\lambda_0\rightarrow 0$ when 
$T_c,\Delta_{1}, \Delta_{2} \rightarrow 0$ 
the gap ratio is not given by the ratio of the densities of states and depends 
on $T_c$ and thus $\lambda_0$.
The ratio $\Delta_1 \over \Delta_2$ decreases from the value much larger than $\sqrt{N_2 \over N_1}$ 
for small $\lambda_0$ to values  below $\sqrt{N_2 \over N_1}$ for larger $\lambda_0$.  
However, it is interesting to note that the correct description of the intra-band 
superconductivity requires the strong coupling theory \cite{dolgov2009}, even if  $\lambda_0<1$ 
and   Van Hove singularity plays similar role in Eliashberg equations \cite{markiewicz1997}.

\subsection{The role of the band couplings}

Before the presentation of the real space local properties of the model 
with general interactions   we spent here 
some time on discussing the homogeneous systems and the influence of model parameters 
on the superconducting bulk state. In particular, we are interested in
the dependence of superconducting state on the couplings between bands. 
The hybridization parameter $t_{12}$ provides single particle coupling and
the inter-band pair scattering $U_{12}$ provides the direct 
two body inter-band interaction. The hybridization  changes single
particle spectrum and this influences the superconductivity.

\begin{figure}
\rotatebox{-90}{\scalebox{0.35}{
\includegraphics{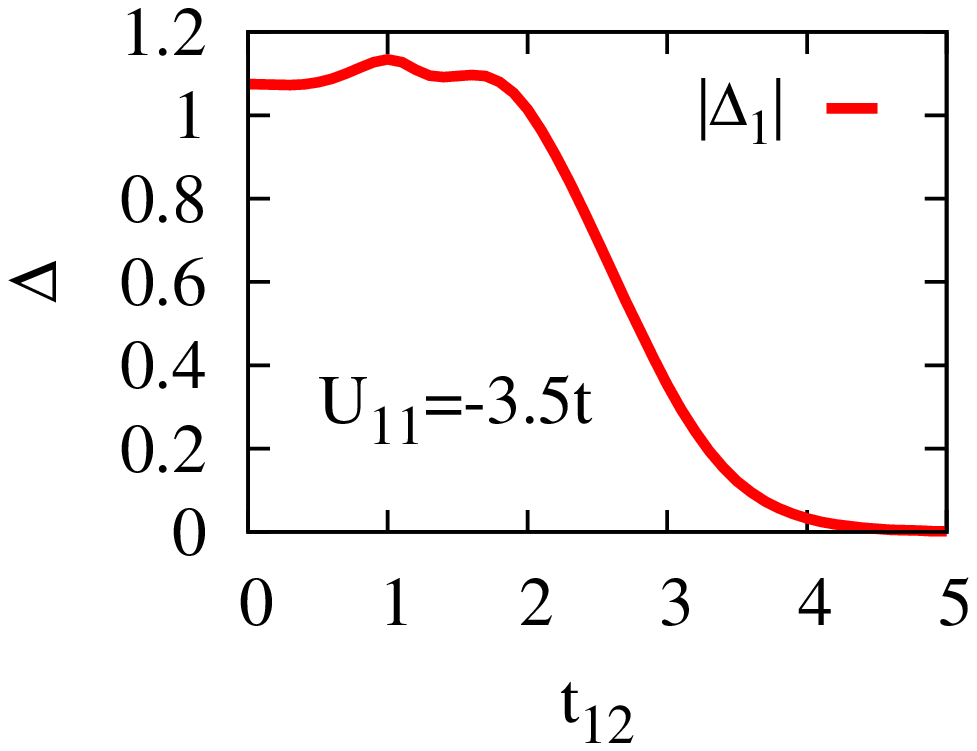}}}
\rotatebox{-90}{\scalebox{0.35}{
\includegraphics{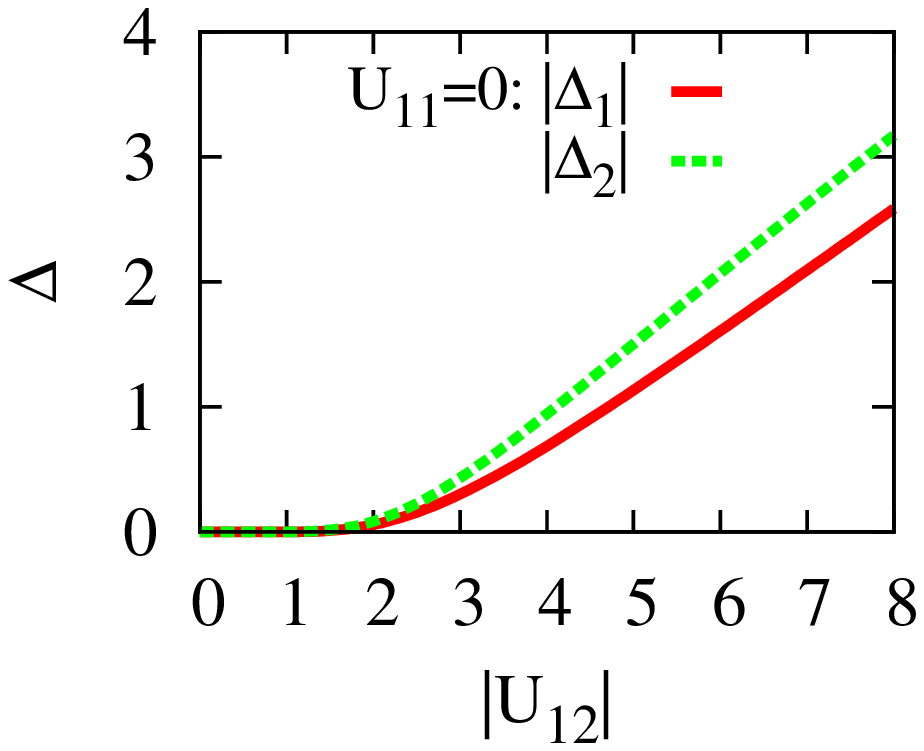}}}
\caption{\label{del-t12} Increase of the hybridization results
in strong decrease of $\Delta_{1}$ for $t_{12}>2$ (left panel). 
For a given set of parameters 
($e_2-e_1=2t, t_1=t,t_2=2t,U_{11}=-3.5t$, $n=1.2$) it is mainly due to
diminishing of the density of states near the Fermi energy (cf. Fig. (\ref{dos-t12})).
Right panel shows the dependence of order parameters in both bands on
the modulus of the intraband interaction $U_{12}$ in a model with $t_{12}=0$ and
$U_{11}=U_{22}=0$.}
\end{figure}

\begin{figure}
\rotatebox{-90}{\scalebox{0.32}{
\includegraphics{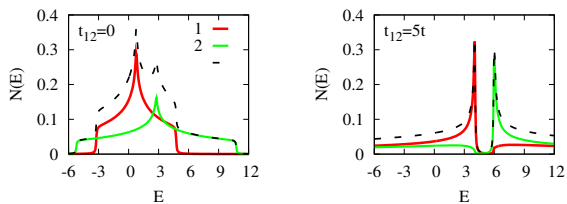}}}
\caption{\label{dos-t12} Changes in the normal   density of states
(total - the dashed curve and projected onto respective orbitals- solid curves) 
 for the system as in the  Fig. (\ref{del-t12}) with the
hybridization $t_{12}=0$ (left panel) and $5t$ (right panel). The Fermi 
energy is at E=0.}
\end{figure}

Fig. (\ref{del-t12}) (left panel) shows the changes of the order parameter in the 
first band  due to increase of hybridization $t_{12}$. The strong  decrease of 
$\Delta_{1}$ with $t_{12}$ results from the changes of the single particle spectrum
in the first band. The second band is essentially decoupled, as $U_{12}=0.$  
One observes strong decrease of the projected density of states around the Fermi 
level in both bands. This is illustrated in  the figure (\ref{dos-t12}). 
  Left panel shows the total (dashed curve) and  projected onto orbital 1 and 2 
densities of states in the system without any inter-band coupling ($t_{12}=0$), 
while in the right panel for strong hybridization $t_{12}=5t$.

As mentioned, the inter-band interaction alone leads to
the superconducting instability independently if it
is repulsive or attractive. It induces gaps in both bands. 
The results are shown in the right panel of Fig. (\ref{del-t12}). The value of the gap 
in the {\it second band} is larger because the density of states near E$_F$ {\it in the first band}
 is larger ({\it c.f.} equations (\ref{intrab})).  

\begin{figure}
\rotatebox{-90}{\scalebox{0.35}{
\includegraphics{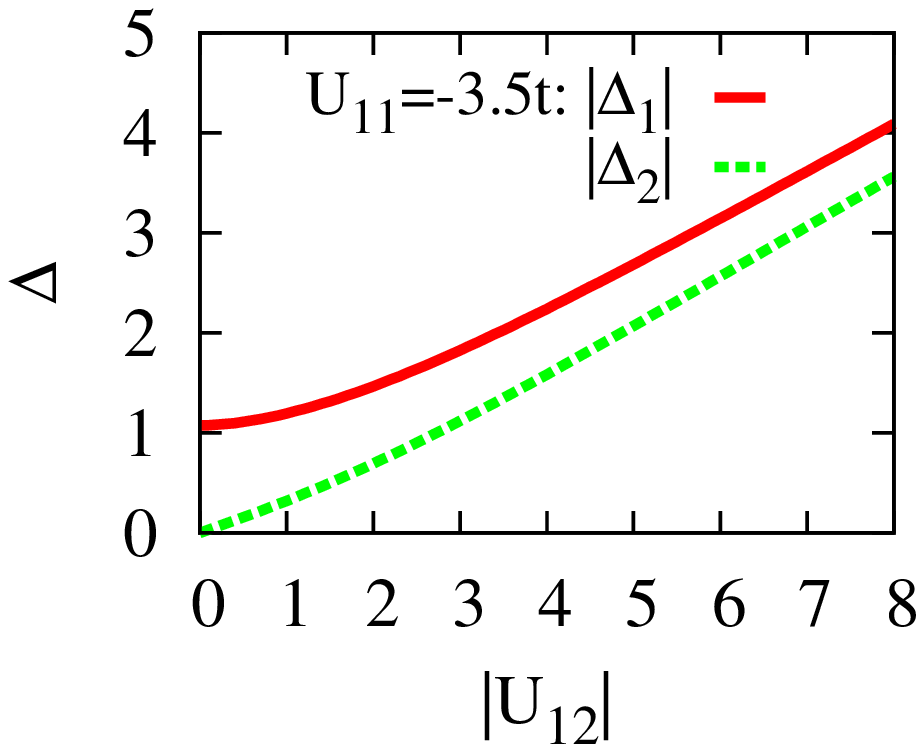}}}
\rotatebox{-90}{\scalebox{0.32}{
\includegraphics{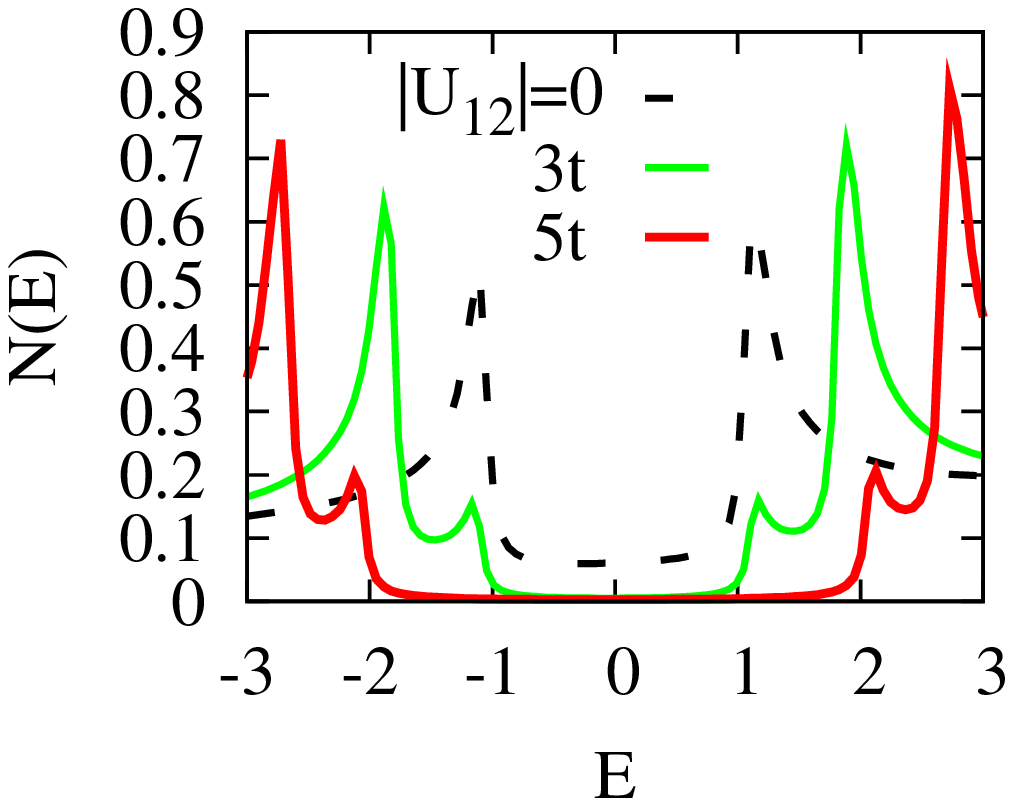}}}
\caption{\label{del-U12+U11}  The dependence of $\Delta_{1}$ and $\Delta_{2}$ 
on $U_{12}$ in the superconductor with intraband attraction $U_{11}=-3.5t$ (left panel). 
The right panel shows the energy dependence of the quasiparticle density of states
for few values of the interband interaction between bare bands ($t_{12}=0$). The structure
in the quasiparticle DOS is due to the gap induced in the non-active band.}
\end{figure}

The simultaneous presence of the inter-band ($U_{12}$) and intra-band (here $U_{11}$ only) 
interactions results in an increase of the order parameter in the active 
band, the appearance of $\Delta$ in the nonactive band and  characteristic 
modifications of the quasiparticle density of states as illustrated in Fig.  
(\ref{del-U12+U11}). The nonzero density of states around chemical potential
for $U_{12}=0$ is simply a result of the absence of couplings between bands 
(as also $t_{12}=0$)
and the lack of pairing in the second band. This presents (slightly artificial)
case of coexisting normal electrons and coherent Cooper pairs in the system.

This ends up our analysis of the homogeneous two band superconductors.
In the next sections we study various inhomogeneities starting with single
potential scatterers.

\section{Single potential impurity in a clean superconductor}
In this section we study a single short-ranged non-magnetic 
impurity embedded in an otherwise clean system.
We solve BdG equations (\ref{BdGEa}) and  (\ref{BdGEb}) on a small cluster
of size $L=13\times 17$ with an impurity placed in its center. 
In the two orbital model the impurity may scatter electron from a given orbital
to the same orbital (intra-orbital scattering to be denoted $V^{1(2)}_{imp}$) 
or to other orbital (inter-orbital scattering - $V^{12}_{imp}$) located at the same site. 
The inter-band scatterers were intensively studied \cite{mitrovic2004} in connection with MgB$_2$.
It has been found that Eliashberg theory leads to much slower rate of T$_c$ suppression
than  predicted on the basis of BCS treatment. 
Here we  allow for both, the intra-band and inter-band pairing interactions
and compare the T$_c$ changes induced by two types of impurities. Instead
of strong coupling Eliashberg approach we are using Bogolubov-de Gennes approach which
allows for the distortion of the wave function around impurity and is more suitable
to treat inhomogeneous superconductors than either BCS or Eliashberg 
(both $\vec k$-space based) theories \cite{ketterson1999}.

\begin{figure}
\rotatebox{-90}{\scalebox{0.5}{
\includegraphics{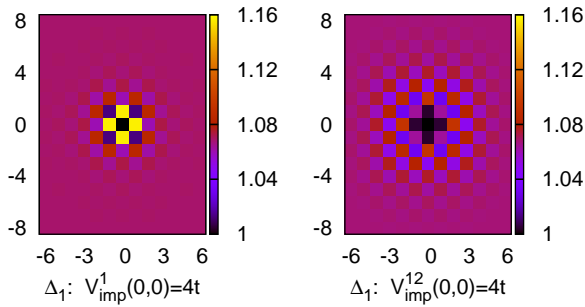}}}
\caption{\label{SI-U1} The influence of intraband $V^{1}_{imp}(\vec{r_i})$ 
(left panel) and interband $V^{12}_{imp}(\vec{r_i})$ (right panel)
 impurity of the same strenght (=4t) on the local values 
of $\Delta_{1}(\vec{r_i})$ in a superconductor
 with  $U_{11}=-3.5t$, $U_{12}=0$.}
\end{figure}

We consider the system described by the following set of parameters  
  $e_2-e_1=2t, t_1=t,t_2=2t,  t_{12}=0.05$ and $n=1.2$, and  
start with  the pairing interaction in the first band band only: 
$U_{11}\ne 0$, $U_{12} = 0$. 
Due to the weak hybridization $t_{12}=0.05t$ there exist small coupling between bands. 
Figure (\ref{SI-U1}) illustrates the changes in the order parameter $\Delta_1$ 
around intra-band (left panel) 
and inter-band (right panel) impurity. Note different patterns of 
changes in $\Delta_1$. The 
intra-band impurity more strongly suppresses order parameter at 
the impurity site and leads to slight 
increase of it (with respect to the value for homogeneous system) 
 at nearest neighbor sites. 
The inter-band impurity scattering on the other hand diminishes the order parameter
at the impurity site and around it, but slightly less for next-nearest neighbors than for
nearest-neighbors. In spite of its short range the inter-band impurity modifies the order
parameter at distances larger than intra-band one.
For the parameters used the clean system has $\Delta_1=1.07t$. 
At the impurity site one finds  $\Delta_{1}(0,0)=0.17t$ for 
$V_{imp}(\vec{r_i})=V^{12}_{imp}(\vec{r_i})$ and $\Delta_{1}(0,0)=0.11t$ for
 $V_{imp}(\vec{r_i})=V^{1}_{imp}(\vec{r_i})$.

\begin{figure}
\rotatebox{0}{\scalebox{0.6}{
\includegraphics{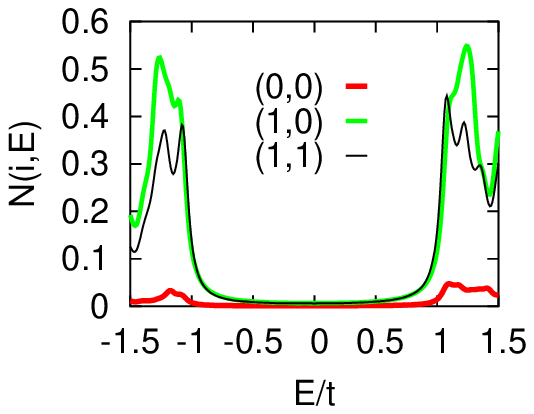}}}
\rotatebox{0}{\scalebox{0.6}{
\includegraphics{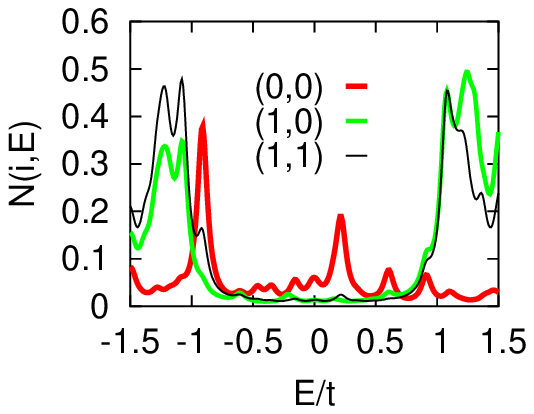}}}
\caption{\label{LDOS-SI-U1} The energy dependence of the local quasiparticle 
density of states projected onto orbital 1,
at the impurity site (0,0) and its neighbours for system with intraband (left panel)
and interband (right panel) impurities for  $U_{11}=-3.5t$, $U_{12}=0$.}
\end{figure}

In figure (\ref{LDOS-SI-U1}) we show the local quasiparticle density of states 
at the impurity site $\vec{r}=(0,0)$ and its nearest $(0,1)$ and next-nearest $(1,1)$ neighbor sites.
The inter-band  impurity induces states inside the gap.  

 The interesting aspect of these studies
 is connected with the fact that the effect of $V^{12}_{imp}$ depends on 
 the sign of inter-band interaction $U_{12}$. There is no similar dependence 
connected with intra-band impurities ($V^1_{imp}$ or $V^2_{imp}$).
 This is illustrated in  the Fig. (\ref{SI-U1,U12,V12}).
 The changes of the order parameters in the first and second bands clearly
 depend on the sign of inter-band interaction.   For the clean system we have $|\Delta_1^0|=2.69t$, $|\Delta_2^0|=2.07t$. At the inter-band impurity site  we have found $|\Delta_1|=1.61t$  
and $|\Delta_2|=1.21t$  for $U_{12}=-5t$, {\it i.e.} roughly 40\% reduction.  
On the other hand in the superconductor with $U_{12}=+5t$ we find 
at the impurity site $|\Delta_1|=0.20t$ and $|\Delta_2|=0.23t$.   The order parameters are  
 suppressed few times stronger in the superconductor with repulsive
inter-orbital interaction.
 
\begin{figure}
\rotatebox{0}{\scalebox{0.6}{
\includegraphics{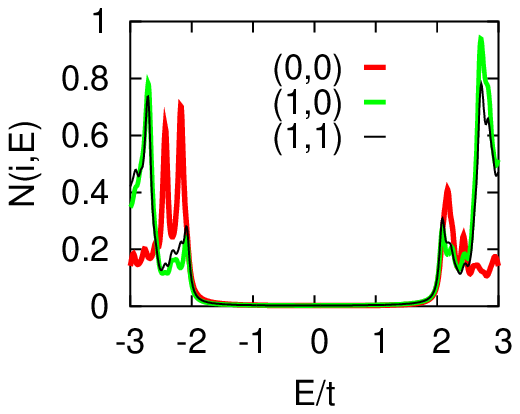}}}
\rotatebox{0}{\scalebox{0.6}{
\includegraphics{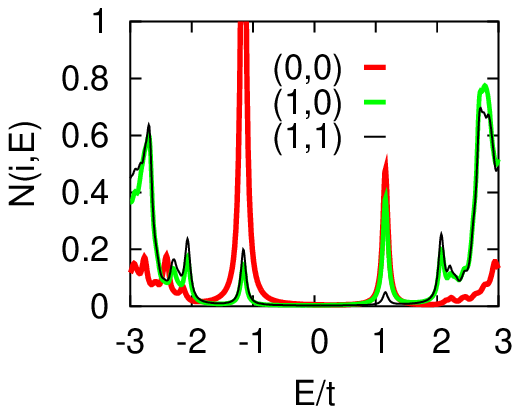}}}
\caption{\label{LDOS-SI-V12U12-5} The energy dependence of the quasiparticle density of states 
at the interband impurity $V_{imp}(\vec{r_i})=V^{12}_{imp}(\vec{r_i})$ and the neighboring sites for
attractive (left panel) and repulsive (right panel) interband pairing interaction. 
}
\end{figure}

  Similar effect has earlier been 
 noted \cite{kogan2009} within the weak coupling Eilenberger theory for finite 
(inter-orbital) impurity concentration in the  
 two band superconductors. Here we observe similar behavior already for
 the single  impurity. The feature is  farther discussed in the next section for
a system with finite concentration of impurities.
 Performing analytical studies of the $T_c$ suppression in two band case the authors \cite{kogan2009}
 have noted that strong suppression of superconductivity for finite 
 concentration of inter-band impurities is to be expected for the inter-band couplings 
 fulfilling the inequality 
 \begin{equation}
 U_{12}\geq -{{N_1^2U_{11}+N_2^2U_{22}} \over{2N_1N_2}}.
 \end{equation}
 It  other words much weaker suppression of $T_c$ is expected for attractive 
 inter-band interaction (note that positive interactions are
 attractive in the notation of the paper [\onlinecite{kogan2009}]). 
 In more detail this is again illustrated 
 in Fig. (\ref{LDOS-SI-V12U12-5}), which shows the energy dependence of the quasiparticle
 density of states at and near the impurity site
for attractive (left panel) and repulsive (right panel)
 inter-band interaction. In the later case the order parameter  
at the impurity site and around it 
is strongly suppressed and new states appear inside the   gap.  
The intra-band impurity scattering (not shown) is not 
effective in suppressing superconductivity 
independently of the $U_{12}$ sign. This is 
due to the same mechanism (Anderson theorem) as for single band
s-wave superconductors \cite{anderson1959}.

\begin{figure}
\rotatebox{-90}{\scalebox{0.33}{
\includegraphics{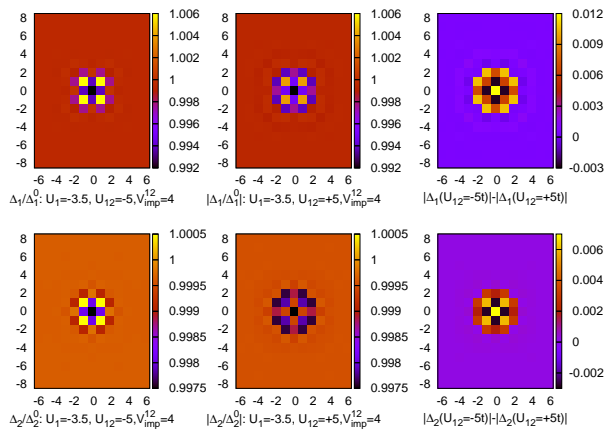}}}
\caption{\label{SI-U1,U12,V12} The changes of the order parameters 
around inter-band impurity  $V^{12}_{imp}$ located at the center 
of the superconducting cluster with $U_{11}=-3.5t$. Left panels are for 
$U_{12}=-5t$, middle panels for  $U_{12}=5t$. The right panels
show the maps of difference between values of the order parameter 
for attractive and repulsive interband interaction.}
\end{figure}

It is also of interest to look at the relative phases of
the order parameters near the impurity in the superconductor with 
$U_{12}>0$. As noted earlier for repulsive inter-orbital interactions 
the $s_{\pm}$-wave pairing state \cite{mazin2008,seo2008} is realised.
This state has s-wave like order parameters on two Fermi surfaces,
the phases of which differ by $\pi$. It means that if the phase of $\Delta_{\lambda}$ 
on one of the Fermi surfaces is $\phi$, the phase on the other is $\phi +\pi$. This phase
relation is in fact responsible for strong suppression of superconductivity 
by inter-band impurities. It turns out that at the $V^{12}_{imp}$ impurity 
site the phases of the gaps change by $\pi$ with respect to the phases in the
bulk. Obviously, there is no such effect for attractive inter-band interaction.

\section{Many impurities in the two band superconductor}
In this section we consider the two band impure superconductor with  
 inter-band and intra-band impurities. Our sample has a rectangular 
shape. It is $L=17\times 21$ sites large with a square lattice of unit lattice constant.
It contains 20\% inter-band or intra-band impurities randomly distributed.
We are not averaging over distribution of impurities, but rather calculate 
the local property at each site and present the result in the form of maps.
To make the impurities more realistic we assume that they are extended
$V^{\lambda\lambda '}_{imp}(\vec{r_i})=V^{\lambda\lambda '}f_{id}$ with 
$f_{id}$ being a number from the Gaussian distribution 
at sites distance $id=1, \sqrt{2}, 2$  from
the impurity. The superconductor studied in this Section is characterized by one 
active band with $U_{11}=-3.5t$ and inter-band interaction $|U_{12}|=5t$. The other
parameters are:  $e_2-e_1=2t, t_1=t,t_2=2t,  t_{12}=0.05$ and $n=1.2$. Energies are 
measured in units of $t$ with respect to chemical potential.

\begin{figure}
\rotatebox{-90}{\scalebox{0.35}{
\includegraphics{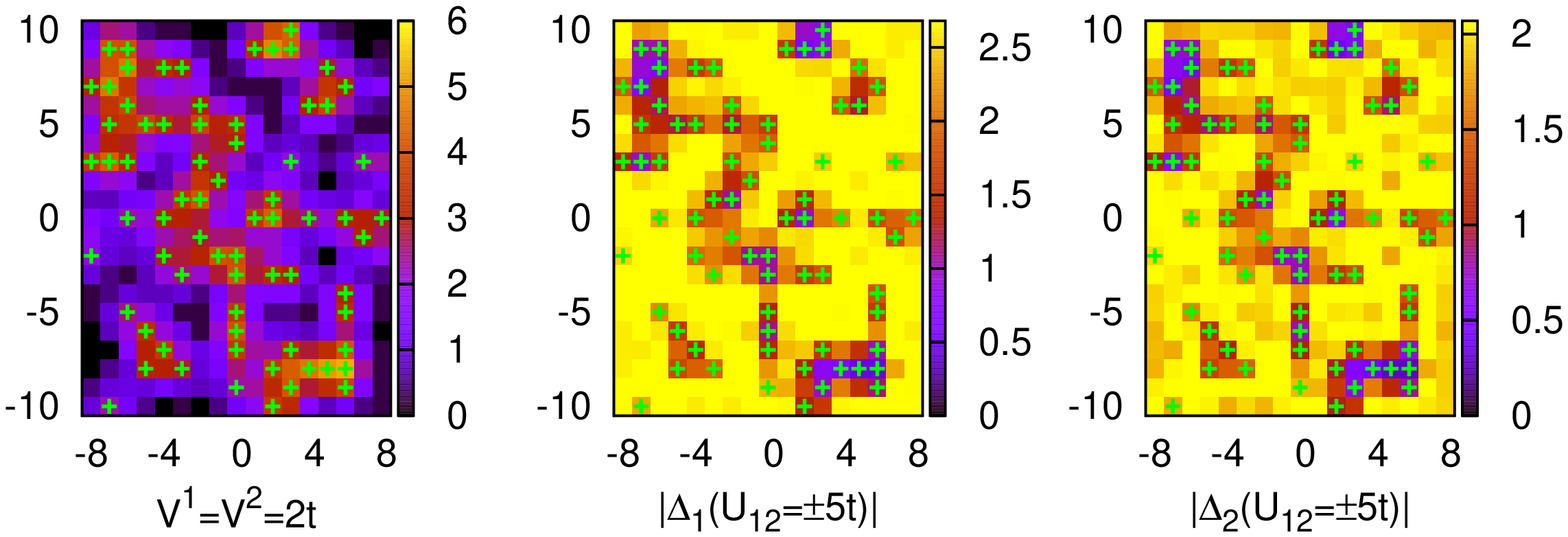}}}
\caption{\label{V1=V2-mapa} Distribution of the intra-band impurity potential 
$V^{1}_{imp}(\vec{r_i})=V^{2}_{imp}(\vec{r_i})$ (left panel), local values of the order parameter 
in the first  (middle panel) and second band  (right panel). The results do not depend 
if the inter-band scattering $U_{12}=\pm 5t$ is of repulsive or attractive type. }
\rotatebox{-90}{\scalebox{0.5}{
\includegraphics{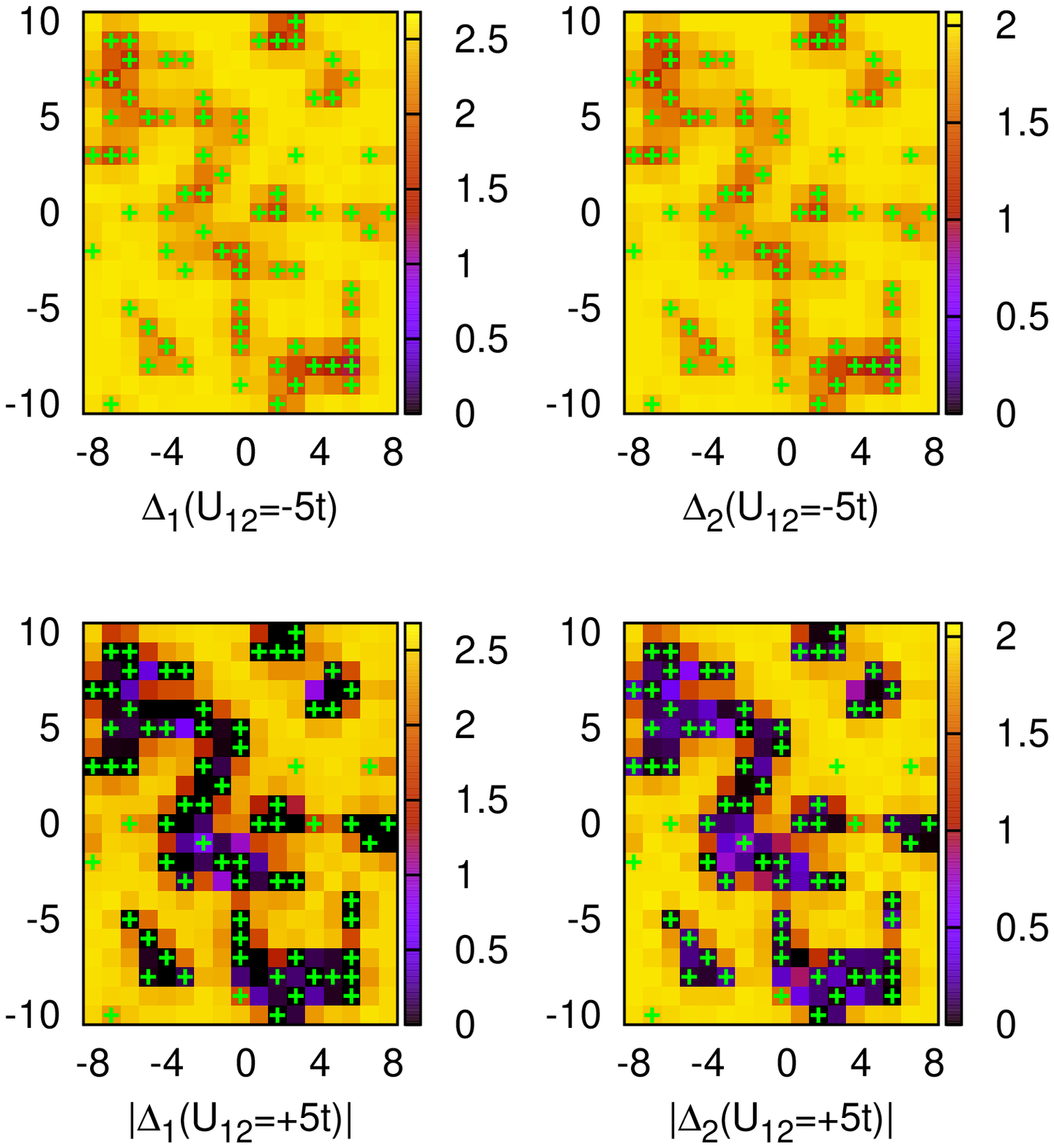}}}
\caption{\label{V12-mapa} The maps of the local values of the 
order parameters in the first band (left panels) and in the second band (right panels).
The upper two figures are obtained for attractive  inter-orbital interaction 
$U_{12}=-5t$, while lower figures represent data for repulsive $U_{12}=5t$ interaction.}
\end{figure}
Figures (\ref{V1=V2-mapa}) and (\ref{V12-mapa}) show the suppression of the
order parameters in the two bands for intra-band and inter-band impurities, respectively.
In all cases one observes similar patterns and large degree of (anti)correlation between the
impurity positions and the gap values. However, the inter-band impurities 
suppress order parameters in  both bands much less for $U_{12}=-5t$ 
than for opposite sign of this  coupling. For negative $U_{12}$ the phases
of the order parameters on two Fermi surfaces are the same and the scattering
of a pair from band 1 into band 2 is harmless, as the superconductor as
a whole looks like s-wave one, and is protected against impurities by the
Anderson theorem \cite{anderson1959}. 

\begin{figure}
\rotatebox{-90}{\scalebox{0.3}{
\includegraphics{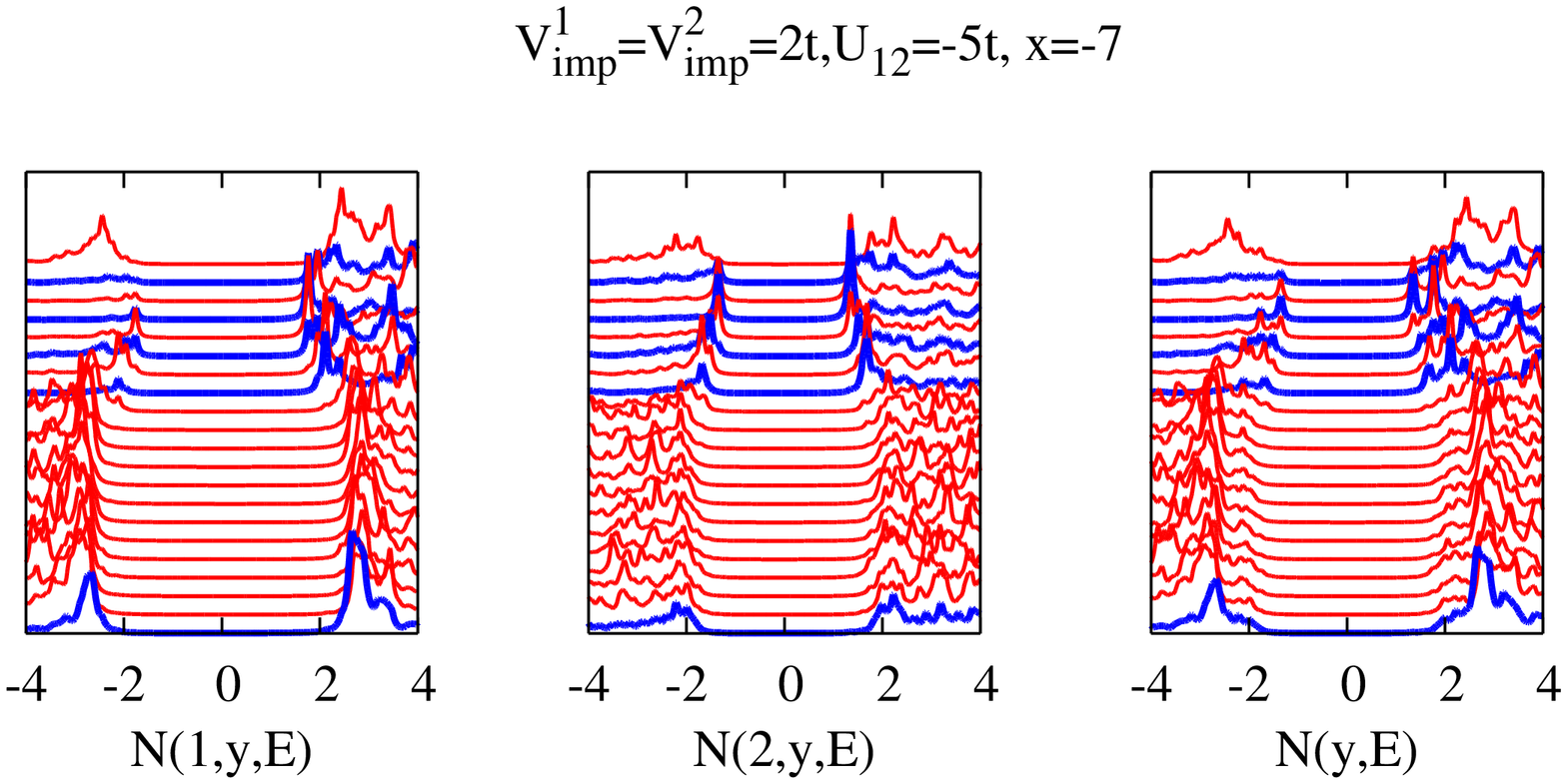}}}
\caption{\label{V1=V2-LDOS} 
Partial (left and middle panel) and total (right panel) 
density of states as a function of energy close to the Fermi energy
at the sites along the line $x=-7$ of the same sample as 
shown in the figure (\ref{V1=V2-mapa}).} 
\rotatebox{-90}{\scalebox{0.3}{
\includegraphics{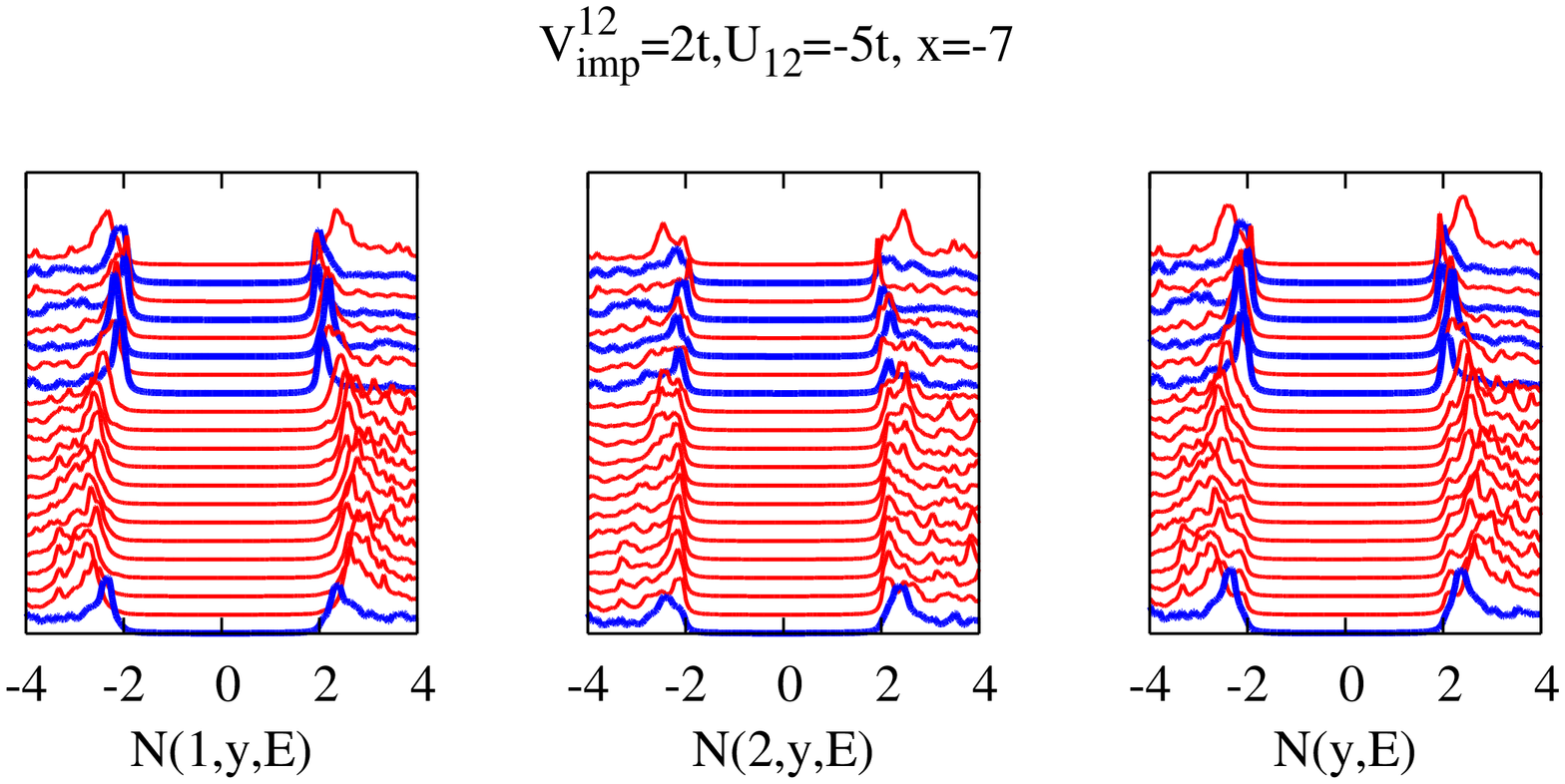}}}
\rotatebox{-90}{\scalebox{0.3}{
\includegraphics{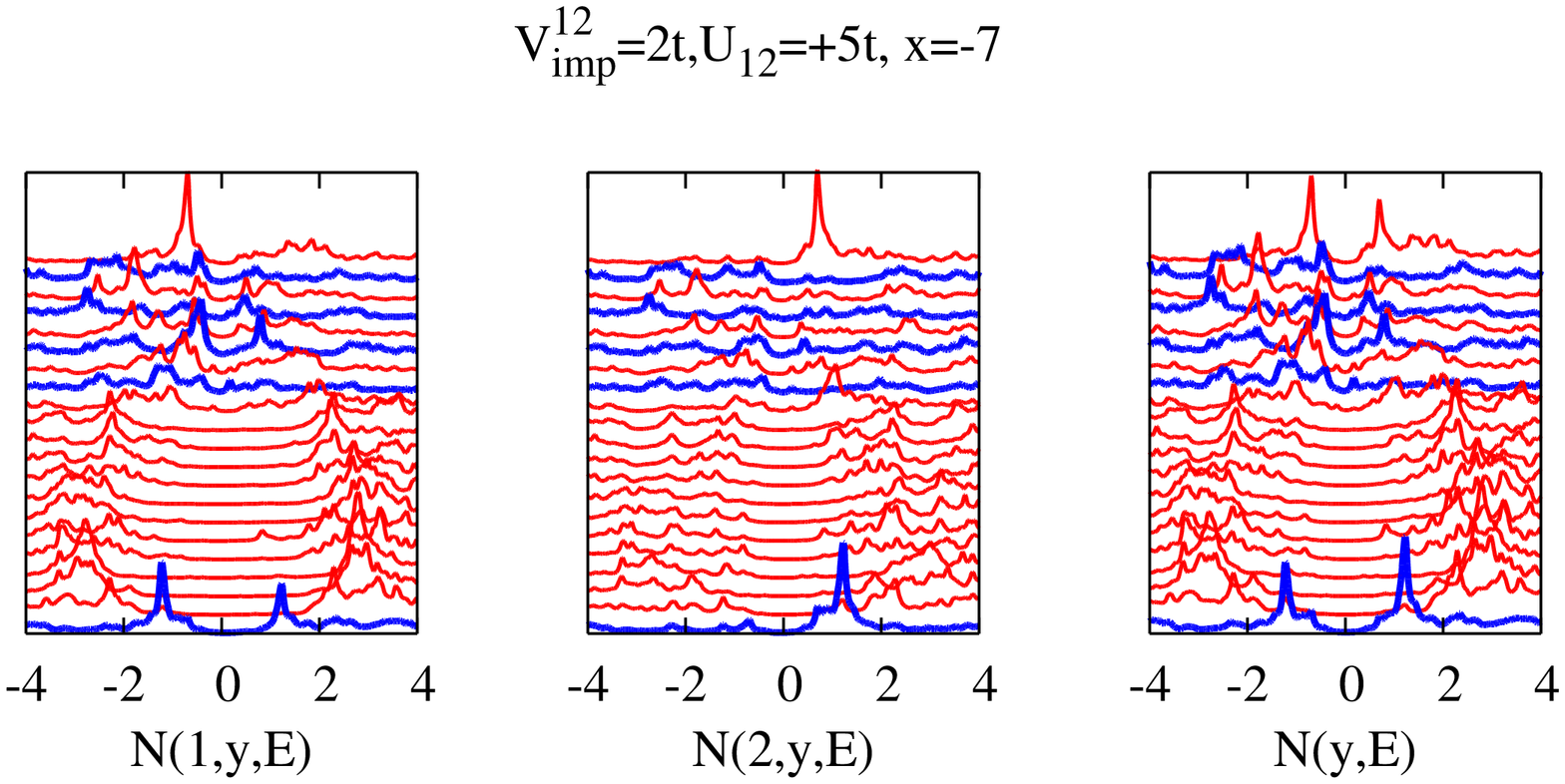}}}
\caption{\label{V12-LDOS} The energy dependence of the 
partial and total densities of states $N(\lambda,y,E)$ at the 
sites along the line $x=-7$ of the sample
shown in figure  (\ref{V12-mapa}). The upper row corresponds to attractive
inter-band interaction, while the lower to repulsive one.}
\end{figure}

Even though the maps presenting suppression of the order parameters 
by the intra-band and inter-band
impurities shown in the Figures (\ref{V1=V2-mapa}) and (\ref{V12-mapa}) 
look to large extend similar the big differences are observed
in the local densities of states. They are shown in figures  
(\ref{V1=V2-LDOS}) and (\ref{V12-LDOS}). Left panels of both figures show the
local densities of states as function of energy 
along the line $x=-7$ in the first band. Middle panels present the results for
 the second band
and the total DOS is plotted in right panels along the same cut. 
Intra-band impurities, Fig. 
(\ref{V1=V2-LDOS}) induce large inhomogeneities, which show up as 
gaps in the local density of states of amplitude strongly changing from site to site.
Sites close in space may have largely different gaps. Similarly, the inter-band
impurities in a superconductor with large attractive inter-band scattering
also induce inhomogeneities. However, they are much smaller (recall for the same distribution
and strength of impurities). The gaps also change from site to site, but
more gradually {\it i.e.} on larger spatial scale.  
On the contrary, the same inter-band impurities nearly completely destroy 
the superconductivity in the system with strong repulsive inter-band
pair scattering {\it i.e.} in the $s_{\pm}$ state.

\begin{figure}
\rotatebox{-90}{\scalebox{0.3}{
\includegraphics{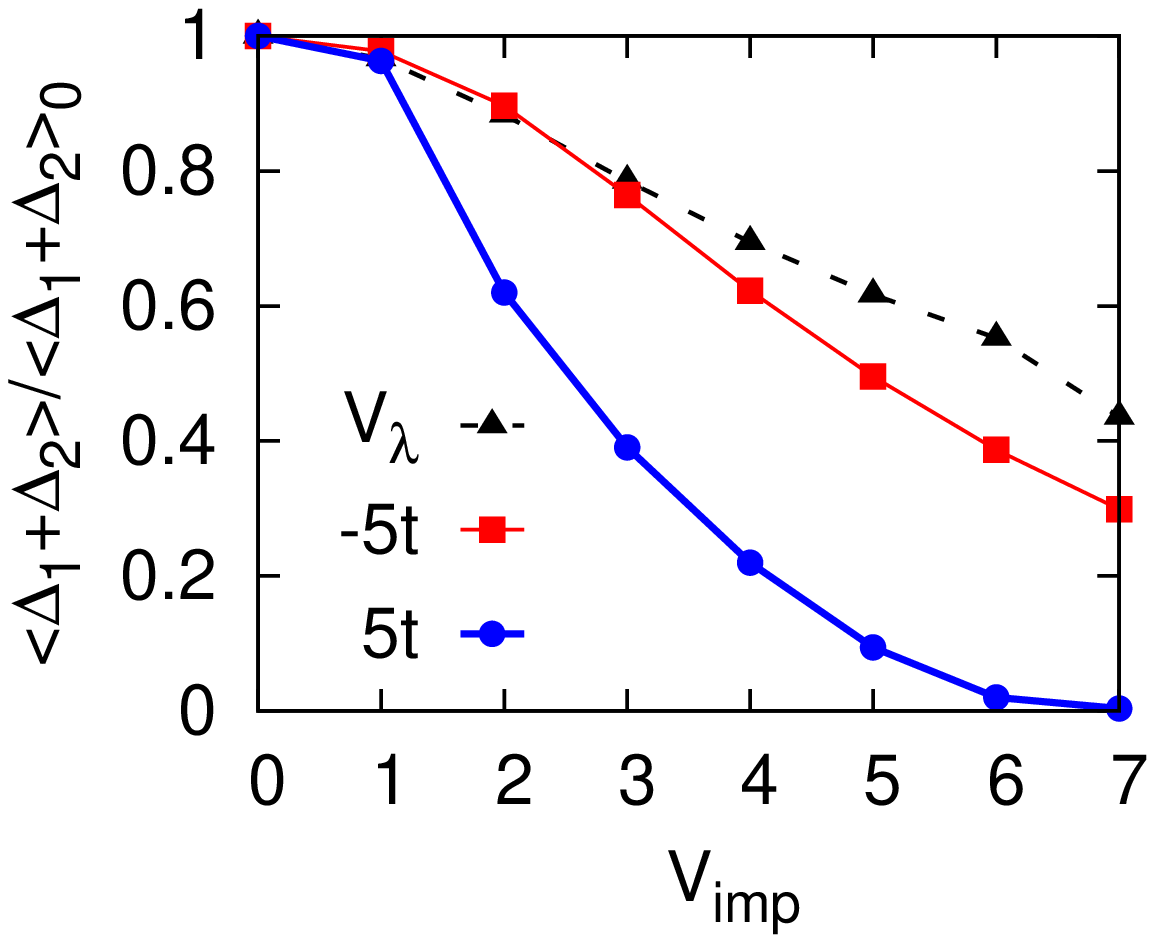}}}
\rotatebox{-90}{\scalebox{0.3}{
\includegraphics{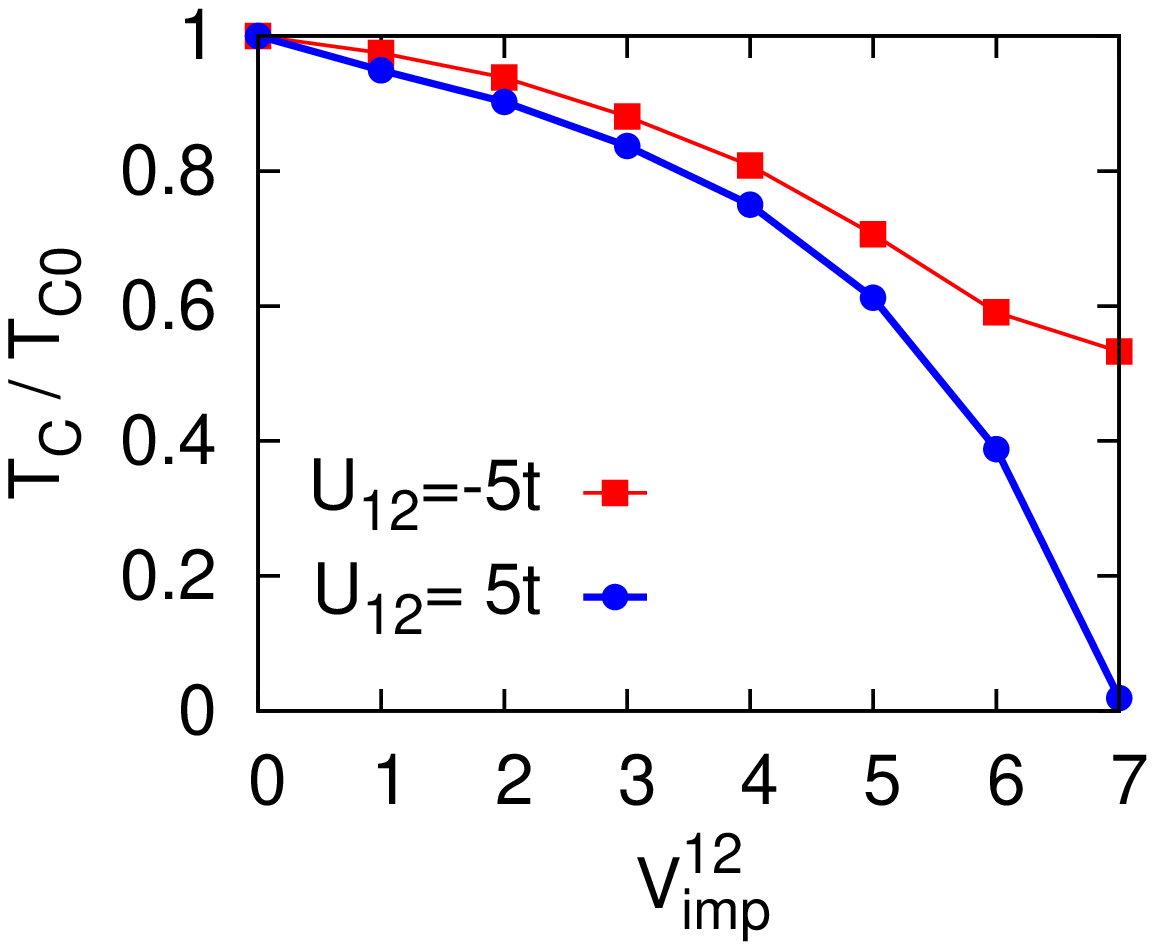}}}
\caption{\label{V12-Tc} The left panel shows the dependence of the normalised
  average order parameter on the $V_{imp}=V^1=V^2$ (dashed curve with triangles) and
on the  $V_{imp}=V^{12}$ with $U_{12}=-5t$ (curve with full squares) or $U_{12}=+ 5t$
 (curve with dots). The right panel  shows the changes in $T_c$ with 
 $V^{12}$ for two signs of inter-band scattering.}
\end{figure}

Different reaction of the system with repulsive and attractive inter-band interactions
to  impurities may well be characterized by the dependence of the
superconducting transition temperature $T_c$  or the average gap 
on the strength of $V_{imp}$. 
In the left panel of the Fig.  (\ref{V12-Tc}) we show the relative change  of 
$<\Delta_{1}+\Delta_{2}>$ in the impure system normalized to its 
clean value $<\Delta_{1}+\Delta_{2}>_0$
 with increasing  the strength of intra-band impurities  
$V_{imp}=V_{imp}^1=V_{imp}^2$ (dashed curve with triangles) and on 
 inter-band impurities  $V_{imp}=V_{imp}^{12}$ in a system with  $U_{12}=- 5t$ 
(curve with full squares) or $U_{12}=+ 5t$ (curve with dots). 
The right panel of that figure shows the changes in $T_c$ with 
 $V_{imp}^{12}$ for two signs of inter-band scattering. In a qualitative agreement with 
the results \cite{kogan2009} 
 of Kogan {\it et al.} we observe much stronger diminishing of $T_c$ normalized 
to its clean system value $T_{c0}$ for repulsive than for attractive $U_{12}$. Similar
  dependence of $T_c$ on impurity strength for both signs of $U_{12}$ 
at small disorder is attributed to our use of Bogolubov - de Gennes
approach which is more suitable to study inhomogeneous systems than the Eilenberger or BCS
theories. In BdG approach the condensate wave function  may distort around the impurity 
and this allows the system to keep its condensation energy and $T_c$ much higher than
it would result from Abrikosov-Gorkov approach to impure superconductors, 
which does not allow for a local changes in the wave function. 


\section{Discussion and conclusions}

We have studied the model of two band superconductor
 with intra-band and inter-band interactions. 
In the present paper we concentrated on some general aspects of the
two band model. In particular we have found that the presence of
Van Hove singularity in the density of states in one of the bands
 leads to strong enhancement of the superconducting transition temperature
 of intra-band only superconductor in the weak coupling, 
{\it  i.e.} for $\lambda_{0}\ll 1$. However, neglecting changes of 
prefactor, the mere increase of effective interband coupling
by Van Hove singularity seem to be not large enough to make
the electron-phonon coupling responsible for superconductivity, 
at least so in LaFeAsO for which the coupling constant matrix 
has been found \cite{boeri2009} 
with $\lambda_{12}=0.093$ and $\lambda_{21}=0.124$.  
Interestingly, our analysis 
suggests that for elevated values of $\lambda_{0}$ the effect of Van Hove singularity
is to diminish $T_c$ in comparison to the system without logarithmic
enhancement of DOS. 
The model with both types of pairing interactions leading 
to s -- or $s_{\pm}$--wave symmetry 
displays number of features similar to that observed in real 
many band materials,
particularly MgB$_2$ and iron pnictides.

MgB$_2$ is a well established superconductor with  two gaps \cite{kortus2007}.
It is believed to have one active band and relatively
weak inter-band coupling.  
Its much slower than predicted by the Abrikosov-Gorkov 
theory \cite{golubov1997} suppression of the 
superconducting transition temperature by impurities \cite{eskildsen2002,kazakov2005} 
may point towards the inter-band character of impurities and
repulsive character of inter-band pair scattering ($U_{12}>0$).
Such a case is illustrated in the upper row of Fig. (\ref{V12-LDOS}).

Another prominent recent example of the many band superconductors 
is  provided by the iron pnictides \cite{ishida2009}. These superconductors 
seem to belong to different class of many band materials  in which
the inter-band interaction is dominant and the order parameter has different signs
on different Fermi surface sheets \cite{mazin2008} - the $s_{\pm}$ state. 
The existing samples are certainly strongly disordered, 
as it can be inferred from large values of 
resistance just above $T_c$. Despite  large disorder they are
superconducting with quite large $T_c$. This suggests s- or $s_{\pm}$-wave like
order parameter. As we have seen the superconductor with dominant attractive $U_{12}$
interactions is quite robust against impurities, both of  intra- and (and even more) 
inter-band type  ({\it c. f.} figures (\ref{V1=V2-LDOS}) and (\ref{V12-LDOS})).   

The robustness of the two band superconductors with 
attractive inter-band interactions to the impurities 
can be traced back to the Anderson theorem. 
On the other hand the $s_{\pm}$ state induced
in two band model by repulsive inter-band interactions is characterized
by large sensitivity to inter-band impurity scattering. These results 
are in agreement with previous studies of similar models \cite{kogan2009,mitrovic2004}.

The maps plotted in the Figs. (\ref{V12-mapa},\ref{V1=V2-mapa}) are
in qualitative agreement with recent scanning tunnelling 
microscopy (STM) studies
for pnictide superconductors \cite{yin2009,masse2009,yin2009b}. In these
papers the relatively small variation of the local gaps have been
observed with the average gap $\bar{\Delta}=6 - 7 meV$ and $2\Delta(0)/k_BT_c\approx 7$
indicating strong coupling superconductivity.  
The detailed analysis of the STM spectra in pnictides will be the subject
of future studies.
  
\begin{acknowledgments}
This work has been partially supported 
by the  Ministry of Science and Education under the grant No.  
N N202 1698 36. 
\end{acknowledgments}


\begin{thebibliography}{99}
\bibitem{orbital} We start with the two orbital description of the system and
consider the single and two particle (in the spin singlet Cooper channel) scattering
between orbitals.
\end{thebibliography}
\end{document}